\documentclass[conference]{IEEEtran}
\usepackage[dvipdfm]{graphicx}
\usepackage[usenames,dvipsnames]{color}
\usepackage{amsmath}%
\usepackage{amsfonts}%
\usepackage{amssymb}%
\usepackage{graphicx}%
\usepackage{verbatim}%
\usepackage{enumerate}%
\usepackage{psfrag}
\usepackage{epsfig}

\begin{document}
\title{Power and Transmission Duration Control for Un-Slotted Cognitive Radio Networks}
\author{
\authorblockN{Ahmed~Ahmedin, Marwa~Ali, Ahmed~Sultan and Mohammed Nafie}
\authorblockA{Wireless Intelligent Networks Center (WINC)\\
Nile University, Cairo, Egypt\\
Email: \{ahmed.ahmedin, marwa.ali\}@nileu.edu.eg, \{asultan, mnafie\}@nileuniversity.edu.eg}}



\maketitle

\begin{abstract}
We consider an unslotted primary channel with alternating on/off
activity and provide a solution to the problem of finding the
optimal secondary transmission power and duration given some
sensing outcome. The goal is to maximize a weighted sum of the
primary and secondary throughput where the weight is determined by
the minimum rate required by the primary terminals. The primary
transmitter sends at a fixed power and a fixed rate. Its on/off
durations follow an exponential distribution. Two sensing schemes
are considered: perfect sensing in which the actual state of the
primary channel is revealed, and soft sensing in which the
secondary transmission power and time are determined based on the
sensing metric directly. We use an upperbound for the secondary
throughput assuming that the secondary receiver tracks the
instantaneous secondary channel state information. The objective
function is non-convex and, hence, the optimal solution is obtained
via exhaustive search. Our results show that an increase in the
overall weighted throughput can be obtained by allowing the
secondary to transmit even when the channel is found to be
busy. For the examined system parameter values, the throughput gain from soft sensing is marginal. Further investigation is needed for assessing the potential of soft sensing.\footnote{This work was supported in part by a grant from the
Egyptian National Telecommunications Regulatory Authority }
\end{abstract}
\section{Introduction}
\label{sec:Introduction}

The current scheme of fixed spectrum allocation poses a
significant obstacle to the objective of expanding the capacity
and coverage of broadband wireless networks. One solution to the
problem of under-utilization caused by static spectrum allocation
is cognitive radio technology. In cognitive radio networks, two
classes of users coexist. The primary users are the classical
licensed users, whereas the cognitive users, also known as the
secondary or unlicensed users, attempt to utilize the resources
unused by the primary users following schemes and protocols
designed to protect the primary network from interference and
service disruption. There are two main scenarios for the
primary-secondary coexistence. The first is the overlay scenario
where the secondary transmitter checks for primary activity before
transmitting. The secondary user utilizes a certain resource, such
as a frequency channel, only when it is unused by the primary
network. The second scenario is the underlay system where
simultaneous transmission is allowed to occur so long as the
interference caused by secondary transmission on the primary
receiving terminals is limited below a certain level determined by
the required primary quality of service.

There is a significant amount of research that pertains to the
determination of the optimal secondary transmission parameters to
meet certain objectives and constraints. The research in this area
has two main flavors. The first takes a physical layer perspective
and focuses on the secondary power control problem given the
channel gains between the primary and secondary transmitters and
receivers. The traffic pattern on the primary channel is typically
not included in this approach save for a primary activity factor
such as in \cite{jafar}. On the other hand, the second line of
research concentrates on primary traffic and seeks to obtain the
optimal time between secondary sensing activities in an unslotted
system, or the optimal decision, whether to sense or transmit, in
a slotted system. Usually under this approach the physical layer
is abstracted and the assumption is made that any two packets
transmitted in the same time/frequency slot are incorrectly
received \cite{opp}, \cite{omar} and \cite {HyoilKim}.

In this paper, we assume knowledge of both primary traffic pattern
and channel gains between a primary and a secondary pair. The
objective is to utilize this knowledge to determine both the
optimal secondary transmission power and time after which the
secondary transmitter needs to cease transmission and sense the
primary channel again to detect primary activity. We allow for
secondary transmission even when the channel is perfectly sensed
to be busy. The objective is to maximize a weighted sum of primary
and secondary rates given the channel gains and the primary
traffic distribution functions for the on/off durations. We
consider two sensing schemes: perfect sensing where the secondary
transmitter knows, through sensing, the actual state of the
primary transmitter. The second scheme is soft sensing, introduced
in \cite{jafar}, where secondary transmission parameters are
determined directly from some sensing metric.

The paper is organized as follows: in section \ref{sec:SysMdl} the
system model is introduced. The optimization problem of maximizing
the weighted sum rates is provided in Section \ref{sec:rates}. In
Section \ref{sec:sim_results}, we provide simulation results.
Section \ref{sec:Conc} concludes the paper.

\section{System Model}
\label{sec:SysMdl} We consider an unslotted primary channel with
alternating on/off primary activity similar to the model employed
in \cite{HyoilKim}. We assume that the probability density
function (pdf) of the duration of the on period is exponential and
is given by:
\begin{eqnarray}
\label{ch_pdf }
  f_{\rm on}(t) &=& \lambda_{\rm on} \exp\left(- \lambda_{\rm on} t\right), \ t\geq 0
\end{eqnarray}
where $\lambda_{\rm on}$ is the reciprocal of the mean on duration
$T_{\rm on}$. Similarly, the pdf of the off duration is:
\begin{eqnarray}
\label{ch_pdf2 }
  f_{\rm off}(t) &=& \lambda_{\rm off} \exp\left(- \lambda_{\rm off} t\right), \ t\geq 0
\end{eqnarray}
and $\lambda_{\rm off}=1/T_{\rm off}$, where $T_{\rm off}$ is the mean of the off duration. The channel utilization factor $u$ is given by
\begin{eqnarray}
\label{ch_utilization}
u &=& \frac{T_{\rm on}}{T_{\rm on}+T_{\rm off}}
\end{eqnarray}
Based on results from renewal theory \cite{Cox}, the probability
that the primary channel is free at time $t^{'}+t$ given that it
is free at time $t^{'}$, is given by:
\begin{eqnarray}
\label{ch_transition_prob_free}
      P^{00}(t) &=& (1-u) + u \exp\left(- \left[ \lambda_{\rm on} + \lambda_{\rm off}\right] t\right)
\end{eqnarray}
Given that the channel is busy at time $t^{'}$, the probability of
being free at $t^{'}+t$, is given by:
\begin{eqnarray}
\label{ch_transition_prob_busy}
      P^{10}(t) &=& (1-u)-(1-u)\exp\left(-\left[\lambda_{\rm on}+\lambda_{\rm off}\right]t \right)
\end{eqnarray}
\begin{figure}[t]
    \centering
        \includegraphics[width=.5\textwidth,height=5.5cm]{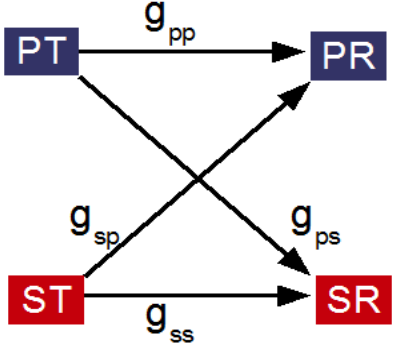}
    \caption{System model where PT denotes the primary transmitter, PR: primary receiver, SR: secondary receiver and ST: secondary transmitter.}
    \label{fig:block}
\end{figure}
The primary transmitter sends with a fixed power $P_{\rm p}$ and
at a fixed rate $r_{\circ}$. A secondary pair tries to communicate
over the same channel utilized by the primary terminals. As seen in Figure
\ref{fig:block}, we denote the gain between primary
transmitter and primary receiver as $g_{\rm pp}$, the gain between secondary
transmitter and secondary receiver as $g_{\rm ss}$, the gain between
primary transmitter and secondary receiver as $g_{\rm ps}$, and
finally the gain between secondary transmitter and primary
receiver as $g_{\rm sp}$. We assume Rayleigh fading channels and, hence, the channel gains are exponentially distributed with mean
values: $\overline{g}_{\rm sp}$, $\overline{g}_{\rm ss}$,
$\overline{g}_{\rm ps}$ and $\overline{g}_{\rm pp}$. The channel
gains are independent of one another, and the primary and
secondary receivers are assumed to know their instantaneous
values. The secondary transmitter does not transmit while sensing the channel. It
senses the channel for a constant time $t_{\rm s}$ assumed to be
much smaller than transmission times ${T}_{\rm on}$ and $T_{\rm
off}$. This assumption guarantees that the primary is highly
unlikely to change state during the sensing period. Based on the
sensing outcome, the secondary transmitter determines its own transmit
power and the duration of transmission after which it has to
sense the primary channel again.
\section{Optimal power level and transmission time }
\label{sec:rates} In this section, we explain the problem of
finding the optimal secondary transmission time and power given
the outcome of the sensing process.
\subsection{Problem Formulation}
\label{prbform} We formulate the cognitive power and transmission
time control problem as an optimization problem with the objective
of maximizing a weighted sum of the primary, $R_{\rm p}$, and
secondary, $R_{\rm s}$, rates. Specifically, we seek to maximize
$\mathbb{E}\{\left(1-\alpha\right)R_{\rm s}+\alpha R_{\rm p}\}$,
where $\mathbb{E}\{.\}$ denotes the expectation operation over the
sensing outcome and primary activity. The constant
$\alpha\in\left[0,1\right]$ is chosen on the basis of the required primary throughput.
The constraints of the optimization problem are
that the secondary power lies in the interval $\left[0,P_{\rm
max}\right]$, and that the time between sensing operations exceeds
$t_{\rm s}$. The problem is generally non-convex and, consequently, we resort to exhaustive search to obtain the solution when the number of optimization parameters is small.

In this paper, we consider two sensing scenarios: 1) perfect
sensing with no sensing errors where the cognitive transmitter knows the exact state of
primary activity after sensing the channel, and 2) soft sensing
where the cognitive transmitter uses some sensing metric $\gamma$,
say the output of an energy detector, to determine its
transmission parameters. Assuming perfect sensing, the parameters
used to maximize the weighted sum throughput are $P_{\rm F}$ and
$T_{\rm F}$ defined as the power and transmission time when the
primary channel is free, and $P_{\rm B}$ and $T_{\rm B}$
corresponding to the busy primary state. Under the soft sensing
mode of operation, the range of values of $\gamma$ is divided into
intervals and the transmission power and time are determined based
on the interval on which the actual sensing metric $\gamma$ lies.
The parameters to optimize the rate objective function are the
transmission powers and times corresponding to each interval and
also the boundaries between intervals.

We assume that the primary link is in outage whenever the primary
rate $r_{\circ}$ exceeds the capacity of the primary channel. The
primary outage probability when the secondary transmitter emits
power $p$ is given by:

\begin{eqnarray}
P_{\circ}\left(p\right)={\rm Pr}\left\{r_{\circ} > \log\left(1+\frac{P_{\rm p}\,g_{\rm pp}}{p\,g_{\rm sp}+\sigma_{\rm p}^2}\right)\right\}
\label{P_outage}
\end{eqnarray}

\noindent where $\sigma_{\rm p}^{2}$ is the noise variance of the
primary receiver. The expression of $P_{\rm o}\left(p\right)$ for
Rayleigh fading channels is given in the Appendix. We assume that
the channel gains vary slowly over time and are almost constant
over several epochs of primary and secondary transmission.

For the secondary rate, we assume that the secondary receiver
tracks the instantaneous capacity of the channel and, hence, the
maximum achievable rate is obtained by averaging over the channel
gains and interference levels [7, equation 8]. The ergodic
capacity of the secondary channel when the cognitive transmitter
emits power $p$ and the primary transmitter is off is expressed as
\begin{eqnarray}
C_{\circ}\left(p\right)=\mathbb{E}_{g_{\rm ss}}\left\{\log\left(1+\frac{p\,g_{\rm ss}}{\sigma_{\rm s}^2}\right)\right\}
\label{sec_cap_no_primary}
\end{eqnarray}
where $\sigma_{\rm p}^{2}$ is the noise variance of the secondary
receiver. When there is simultaneous primary and secondary
transmissions, the ergodic capacity of the secondary channel
becomes
\begin{eqnarray}
C_{\rm 1}\left(p\right)=\mathbb{E}_{g_{\rm ss},g_{\rm ps}}\left\{\log\left(1+\frac{p\,g_{\rm ss}}{P_{\rm p}g_{\rm ps}+\sigma_{\rm s}^2}\right)\right\}
\label{sec_cap_with_primary}
\end{eqnarray}
We provide expressions for $C_{\circ}\left(p\right)$ and $C_{\rm 1}\left(p\right)$ in the Appendix.


\subsection{Perfect Sensing}
\label{sec:hd}
 We mean by perfect sensing that the state of the channel, whether vacant or occupied, is known without error after the channel is sensed. There are four parameters that are used to maximize the weighted sum rate. These are: the secondary power when the channel is sensed to be free, $P_{\rm F}$, the duration of transmission when the channel is sensed to be free, $T_{\rm F}$, the secondary power when the channel is sensed to be busy, $P_{\rm B}$, and the duration of transmission when the channel is sensed to be busy, $T_{\rm B}$. Before formulating the optimization problem under perfect sensing, we need to introduce several parameters that pertain to the primary traffic. The probability, $\pi_{m}$, that the $m$th observation of the channel occurs when the channel is free can be calculated using Markovian property of the traffic model.
\begin{equation}
\pi_{m}=\pi_{m-1} P^{\rm 00}\left(t_{\rm s}+T_{\rm F}\right)+\left(1-\pi_{m-1}\right) P^{\rm 10}\left(t_{\rm s}+T_{\rm B}\right)
\label{pi_m}
\end{equation}
Another parameter is $P^{\rm ss}$ which is the steady state fraction of time the channel is free when sensed according to some scheme. In the perfect sensing scheme, the channel, when sensed free, is sensed again after $t_{\rm s}+T_{\rm F}$. When sensed busy, it is sensed again after $t_{\rm s}+T_{\rm B}$.  Parameter $P^{\rm ss}$ can be obtained by setting $\pi_{m}=\pi_{m-1}=P^{\rm ss}$ in (\ref{pi_m}) to get
\begin{equation}
\label{PSS}
    P^{\rm ss} = \frac{P^{\rm 01}(t_{\rm s}+T_{\rm B})}{1-P^{\rm 11}(t_{\rm s}+T_{\rm F})+ P^{\rm 01}(t_{\rm s}+T_{\rm B})}
\end{equation}
The average time between sensing times is given by:
\begin{eqnarray}\label{Avg}
    \mu &=& P^{\rm ss}\left(t_{\rm s}+T_{\rm F}\right) + (1-P^{\rm ss}) \left(t_{\rm s}+T_{\rm B}\right)
\end{eqnarray}
Finally, we also need the average time the channel is free during a period of $t$ units of time if sensed to be free. We denote this quantity by $\delta^{\circ}\left(t\right)$ and is given by \cite{omar}
\begin{eqnarray}
\label{deltas}
\delta^{\circ}\left(t\right)=\left(1-u\right)\left(t+\frac{ \exp\left[-\left(\lambda_{\rm on}+\lambda_{\rm off}\right)t\right] - 1}{ \lambda_{\rm on}+\lambda_{\rm off} }\right)
\end{eqnarray}
On the other hand, if the channel is sensed to be busy, the average time the channel is free during a period of $t$ units of time is given by \cite{omar}
\begin{eqnarray}
\delta^1(t) & = &   t-u  \left(t+ \frac{ \exp\left[-\left(\lambda_{\rm on}+\lambda_{\rm off}\right)t\right] - 1}{ \lambda_{\rm on}+ \lambda_{\rm off} }\right)
\end{eqnarray}
The secondary throughput averaged over primary activity is given by
\begin{eqnarray}
\overline{R}_{\rm s}&=&P^{\rm ss}\frac{\delta^{\circ}\left(T_{\rm F}\right)}{\mu}C_{\circ}\left(P_{\rm F}\right)+ \nonumber\\
&& P^{\rm ss}\frac{T_{\rm F}-\delta^{\circ}\left(T_{\rm F}\right)}{\mu}C_{\rm 1}\left(P_{\rm F}\right)+ \nonumber\\
&& \left(1-P^{\rm ss}\right)\frac{\delta^{\rm 1}\left(T_{\rm B}\right)}{\mu}C_{\circ}\left(P_{\rm B}\right)+ \nonumber\\
&& \left(1-P^{\rm ss}\right)\frac{T_{\rm B}-\delta^{\rm 1}\left(T_{\rm B}\right)}{\mu}C_{\rm 1}\left(P_{\rm B}\right)
\label{sec_throughput_perfect}
\end{eqnarray}
The first two terms in the above expression are the secondary throughput obtained if the primary is inactive when the channel is sensed. When the sensing outcome is that the channel is free, the secondary emits power $P_{\rm F}$ for a duration $T_{\rm F}$. During the secondary transmission period, the primary transmitter may resume activity. The average amount of time the primary remains idle during a period of length $T_{\rm F}$ after the channel is sensed to be free is obtained by using $t=T_{\rm F}$ in (\ref{deltas}). This is the duration of secondary transmission free from interference from the primary transmitter. On the other hand, the primary transmits during secondary operation for an average period of $T_{\rm F}-\delta^{\circ}\left(T_{\rm F}\right)$. The last two terms in (\ref{sec_throughput_perfect}) are the same as the first two but when the channel is sensed to be busy. In this case, the transmit secondary power is $P_{\rm B}$ and the transmission time is $T_{\rm B}$, of which a duration of $\delta^{\rm 1}\left(T_{\rm B}\right)$ is free, on average, from primary interference.

The primary throughput is given by
\begin{eqnarray}
\label{primary_throughput}
\overline{R}_{\rm p}&=&r_{\circ}P^{\rm ss}\frac{T_{\rm F}-\delta^{\circ}\left(T_{\rm F}\right)}{\mu}\left[1-P_{\circ}\left(P_{\rm F}\right)\right]+ \nonumber\\
&& r_{\circ}\left(1-P^{\rm ss}\right)\frac{T_{\rm B}-\delta^{\rm 1}\left(T_{\rm B}\right)}{\mu}\left[1-P_{\circ}\left(P_{\rm B}\right)\right]
\end{eqnarray}
We ignore the primary throughput that may be achieved during the sensing period because $t_{\rm s}$ is assumed to be much smaller than $T_{\rm on}$and $T_{\rm off}$. The two terms of (\ref{primary_throughput}) correspond to the sensing outcomes of the channel being free and busy, respectively. The optimization problem can then be written as
\begin{eqnarray}
\label{HD_form}
    & & \text{Find: } T_{\rm F}, T_{\rm B}, P_{\rm F} \text{ and } P_{\rm B}\nonumber\\
    & & \text{That maximize: }\left(1-\alpha\right)\ \overline{R}_{\rm s}(T_{\rm F},T_{\rm B},P_{\rm F},P_{\rm B})\nonumber\\
    & & \qquad \qquad \qquad +  \alpha \  \overline{R}_{\rm p}(T_{\rm F},T_{\rm B},P_{\rm F},P_{\rm B}) \nonumber\\
    & & \text{Subject to: } T_{\rm F} \geq 0,\  T_{\rm B} \geq 0,\nonumber\\
    & & 0\leq P_{\rm F} \leq P_{\rm max }\text{ and } 0\leq P_{\rm B} \leq P_{\rm max }\nonumber
\end{eqnarray}

\subsection{Soft Sensing}
\label{sec:sd}
Soft sensing means that the sensing metric is used directly to determine the secondary transmission power and duration. In the sequel, we re-formulate the weighted sum throughput optimization problem assuming quantized soft sensing, where the sensing metric, from a matched filter or an energy detector for instance, is quantized before determining the power and duration of transmission. Let $\gamma$ be the sensing metric with the known conditional pdfs:  $f_{\circ}(\gamma)$ given that the primary is in the idle state and $f_{\rm 1}(\gamma)$ conditioned on the primary transmitter being active. We assume that the number of quantization levels is $S+1$. The $k$th level extends from threshold $\gamma^{\rm th}_{k-1}$ to $\gamma^{\rm th}_{k}$ assuming that $\gamma_{0}^{\rm th}=0$ and  $\gamma^{\rm th}_{S+1}=\infty$. The probability that the metric $\gamma$ is between $\gamma _{k-1}^{\rm th}$ and $\gamma_{k}^{\rm th}$ when the primary channel is free is given by
\begin {eqnarray}
\label{cdf_off}
\epsilon_k&=& {\rm Pr}\{\gamma_{k-1}^{\rm th}\leq\gamma\leq\gamma_{k}^{\rm th}|{\rm channel\ is\ free}\}\nonumber\\
&=& \int_{\gamma_{k-1}^{\rm th}}^{\gamma_{k}^{\rm th}} f_0(\gamma)\ d\gamma
\end{eqnarray}
where $k=1,2,\cdots (S+1)$. On the other hand, The probability that $\gamma$ is between $\gamma _{k-1}^{\rm th}$ and $\gamma_{k}^{\rm th}$ when the primary channel is busy is given by
\begin {eqnarray}
\label{cdf_on}
\vartheta_k&=& {\rm Pr}\{\gamma_{k-1}^{\rm th}\leq\gamma\leq\gamma_{k}^{\rm th}|{\rm channel\ is\ busy}\}\nonumber\\
&=& \int_{\gamma_{k-1}^{\rm th}}^{\gamma_{k}^{\rm th}} f_1(\gamma)\ d\gamma
\end{eqnarray}
When $\gamma$ is between $\gamma _{k-1}^{\rm th}$ and $\gamma_{k}^{\rm th}$, the secondary transmitted power is $P_{k}$ and the duration of transmission is $T_{k}$.

As in the perfect sensing case, the probability that $m$th observation of the channel happens when the channel is free, denoted by $\pi_{\rm m}$, can be calculated using Markovian property of the channel model.
\begin{eqnarray}
\label{probability_sample_k}
\pi _{m}&=&\pi_{m-1} \sum^{S+1}_{k=1} \epsilon_k P^{\rm 00}(t_{\rm s}+T_k)\nonumber\\
&&+(1-\pi_{m-1})\sum^{S+1}_{k=1}\vartheta_k P^{\rm 10}(t_{\rm s}+T_k)
\end{eqnarray}
At steady state, $\pi_{m-1}=\pi_{m}$ and the steady state probability of sensing the channel while it is free becomes
\begin{eqnarray}
\label{pss}
P^{\rm ss}=\frac{\sum^{S+1}_{k=1}\vartheta_k P^{\rm 10}(t_{\rm s}+T_k)}{1-\sum^{S+1}_{k=1}\epsilon_k P^{\rm 00}(t_{\rm s}+T_k)+\sum^{S+1}_{k=1}\vartheta_k P^{\rm 10}(t_{\rm s}+T_k)}
\end{eqnarray}
The average time between sensing events is given by
\begin {eqnarray}
\label{sensing_interval}
\mu=P^{\rm ss} \sum_{k=1}^{S+1}\epsilon_k \left(t_{\rm s}+T_k\right)+(1-P^{\rm ss})\sum_{k=1}^{S+1}\vartheta_k \left(t_{\rm s}+T_k\right)
\end{eqnarray}

The mean secondary throughput averaged over the primary activity and the sensing metric is given by
\begin{eqnarray}
\label{rs_soft}
\overline{R}_{\rm s}&=& P^{\rm ss}\sum^{S+1}_{k=1}\epsilon_k \left[\frac{\delta^{\circ}(T_k)}{\mu}C_{\circ}\left(P_k\right)+\frac{T_k-\delta^{\circ}(T_k)}{\mu}C_{\rm 1}\left(P_k\right)\right]\nonumber\\
&&+(1-P^{\rm ss})\sum^{S+1}_{k=1}\vartheta_k{\Big [}\frac{\delta^{\rm 1}(T_k)}{\mu}C_{\circ}\left(P_k\right)\nonumber\\
&&+\frac{T_k-\delta^{\rm 1}(T_k)}{\mu}C_{\rm 1}\left(P_k\right){\Big ]}
\end{eqnarray}
The mean primary throughput is
\begin{eqnarray}
\label{rp_soft}
\overline{R}_{\rm p}&=& r_{\circ} P^{\rm ss}\sum^{S+1}_{k=1}\epsilon_k \frac{T_k-\delta^{\circ}(T_k)}{\mu}\left[1-P_{\circ}\left(P_k\right)\right]+\nonumber\\
&&r_o(1-P^{ss})\sum^{S+1}_{k=1}\vartheta_k \frac{T_k-\delta^1(T_k)}{\mu}\left[1-P_{\circ}\left(P_k\right)\right]
\end{eqnarray}

\section{Numerical Results}
\label{sec:sim_results}
In this section we present simulation results for the perfect and soft sensing schemes discussed in Section \ref{sec:rates}. The weighted sum rate maximization problem is non-convex, hence, we do exhaustive search to obtain the optimal parameters. The parameters used in our simulations presented here are: $T_{\rm on}=4$, $T_{\rm off}=5$, $t_{\rm s}=0.05$, $r_{\circ}=4.5$ nats, $\sigma_{\rm s}^2=\sigma_{\rm p}^2=1$, $P_{\rm p}=100$, $P_{\rm max}=10$, $\overline{g}_{\rm ss}=2$, $\overline{g}_{\rm pp}=3$, and $\overline{g}_{\rm ps}=.03$. In order to do the exhaustive search, we have imposed an artificial upperbound on transmission time equal to $20$. We analyze the results for perfect sensing in Subsection \ref{sim_hd} and for soft sensing in Subsection \ref{sim_soft}. The parameters for channels \textbf{A} and \textbf{B} used in the analysis are the same except for $\overline{g}_{\rm sp}$ which is equal to $2$ for channel \textbf{A} and $0.2$ for channel \textbf{B}.
\subsection{Perfect Sensing}
\label{sim_hd}
\begin{figure}[t]
    \centering
   \psfrag{R}[][]{$\rm {\alpha\ \overline{R}_p + (1-\alpha)\ \overline{R}_s}$} 
    \epsfig{file=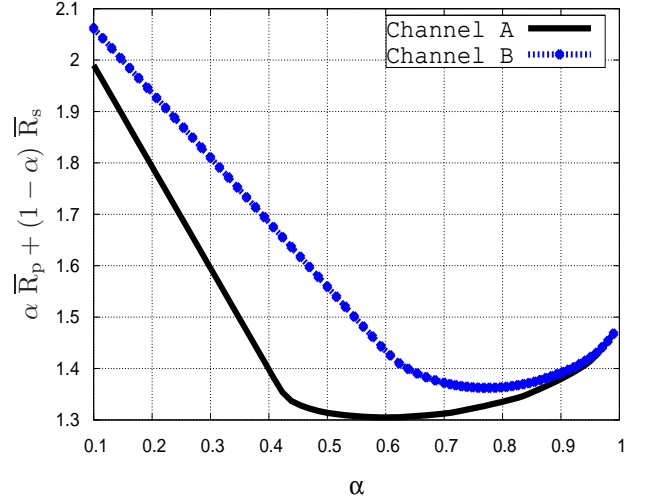, width=0.5\textwidth,height=7cm}
    \caption{Perfect sensing weighted sum throughput versus $\alpha$ for channels \textbf{A} and \textbf{B}.}
    \label{fig:rate_alpha_perfect}
\end{figure}
\begin{figure*}[htpb]
    \centering
    \begin{tabular}{cc}
        \includegraphics[totalheight=.2\textheight,width=.5\textwidth]{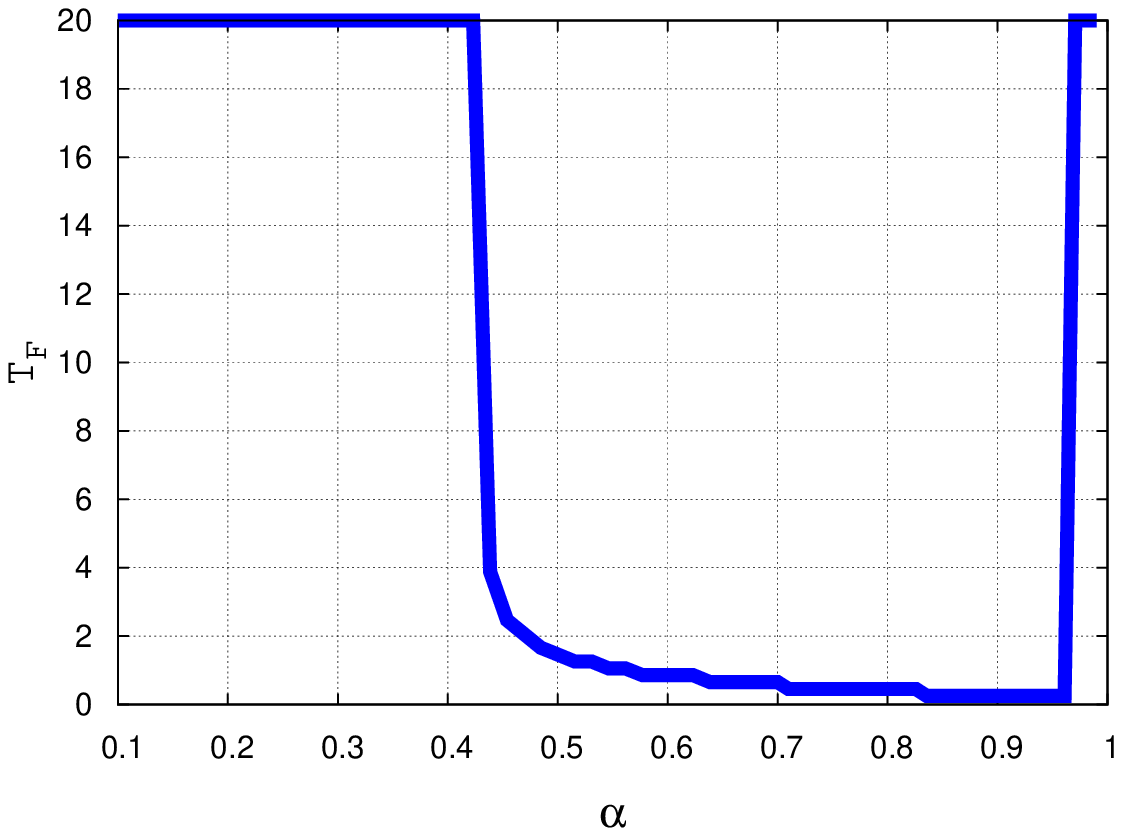}&
        \includegraphics[totalheight=.2 \textheight, width=.5\textwidth]{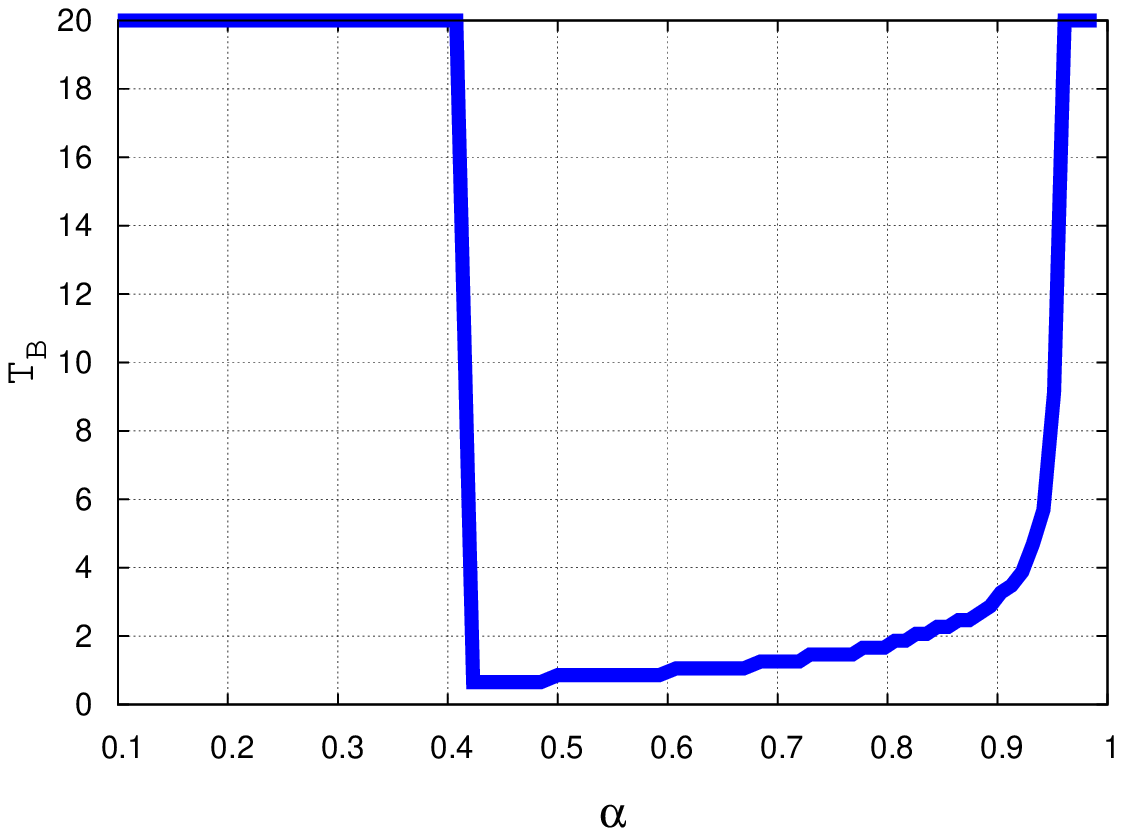}\\
    \includegraphics[totalheight=.2 \textheight, width=.5\textwidth]{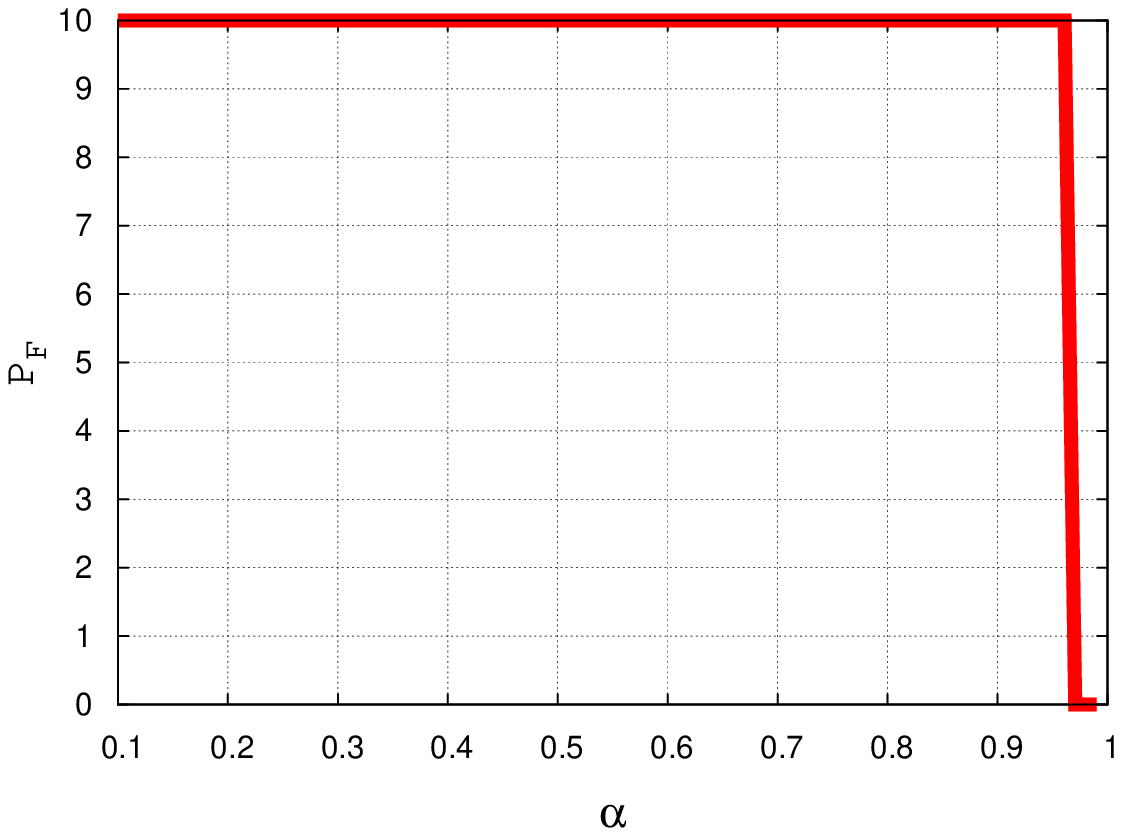}&
    \includegraphics[totalheight=.2 \textheight, width=.5\textwidth]{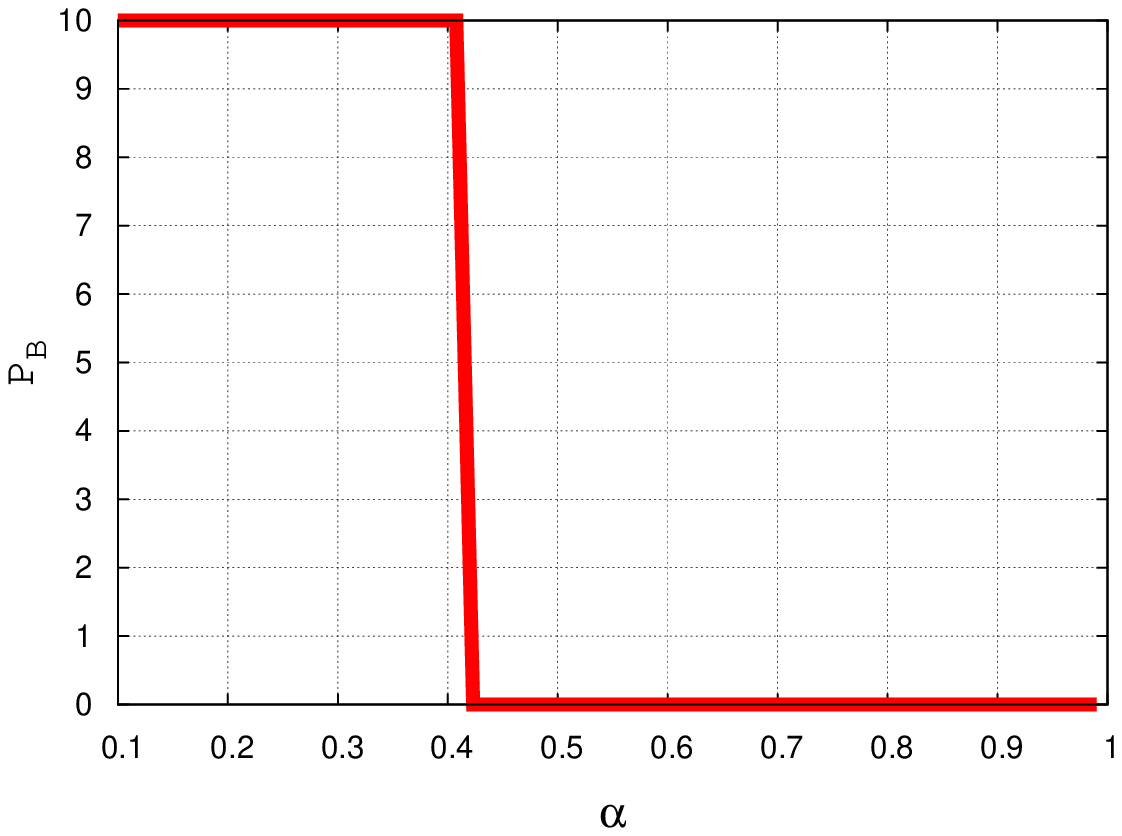}\\
    \end{tabular}
    \caption{Perfect sensing power and transmission time results for channel \textbf{A}.}
    \label{fig:parameter_d_perfect}
%
    \begin{tabular}{cc}
        \includegraphics[totalheight=.2\textheight,width=.5\textwidth]{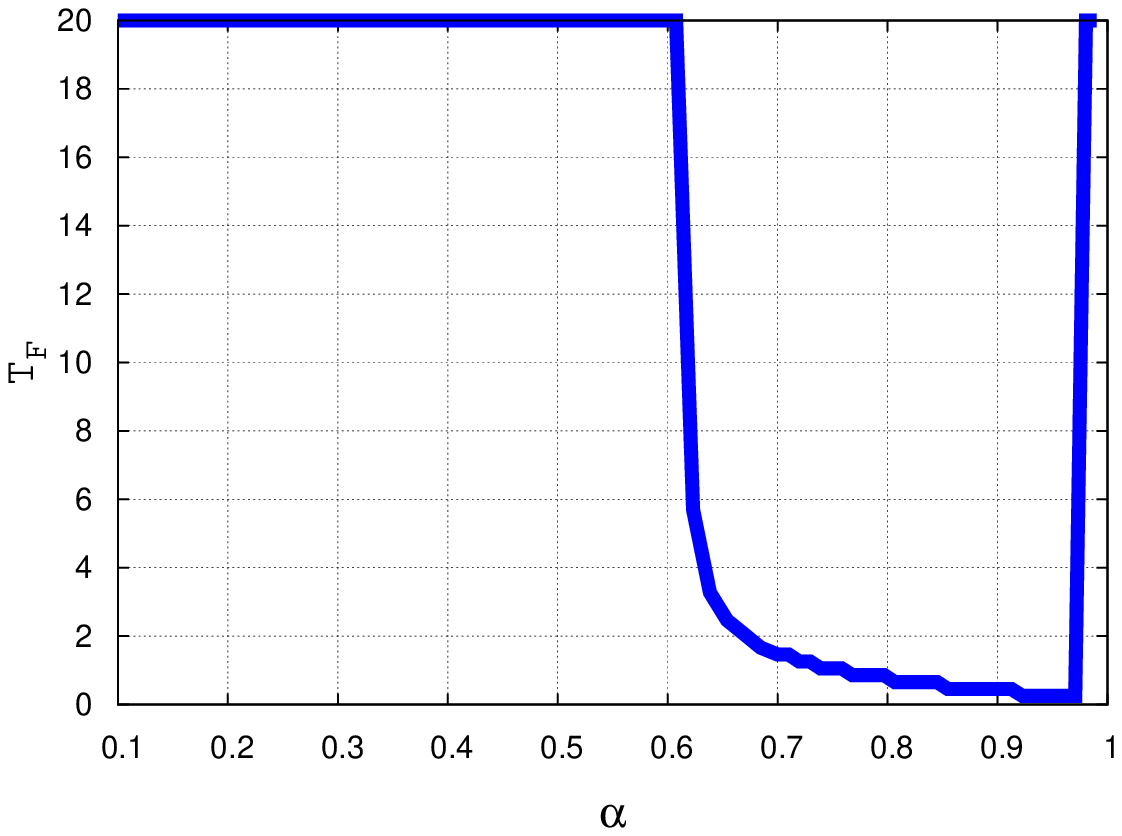}&
        \includegraphics[totalheight=.2 \textheight, width=.5\textwidth]{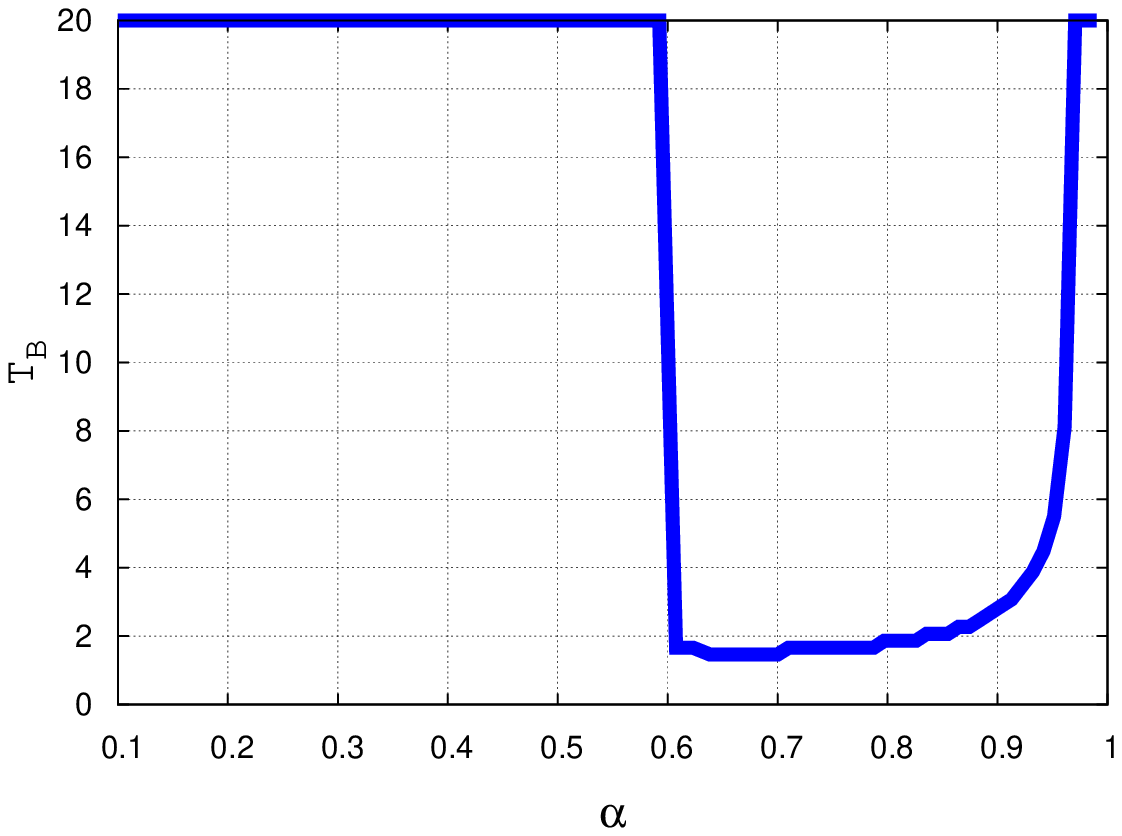}\\
    \includegraphics[totalheight=.2\textheight, width=.5\textwidth]{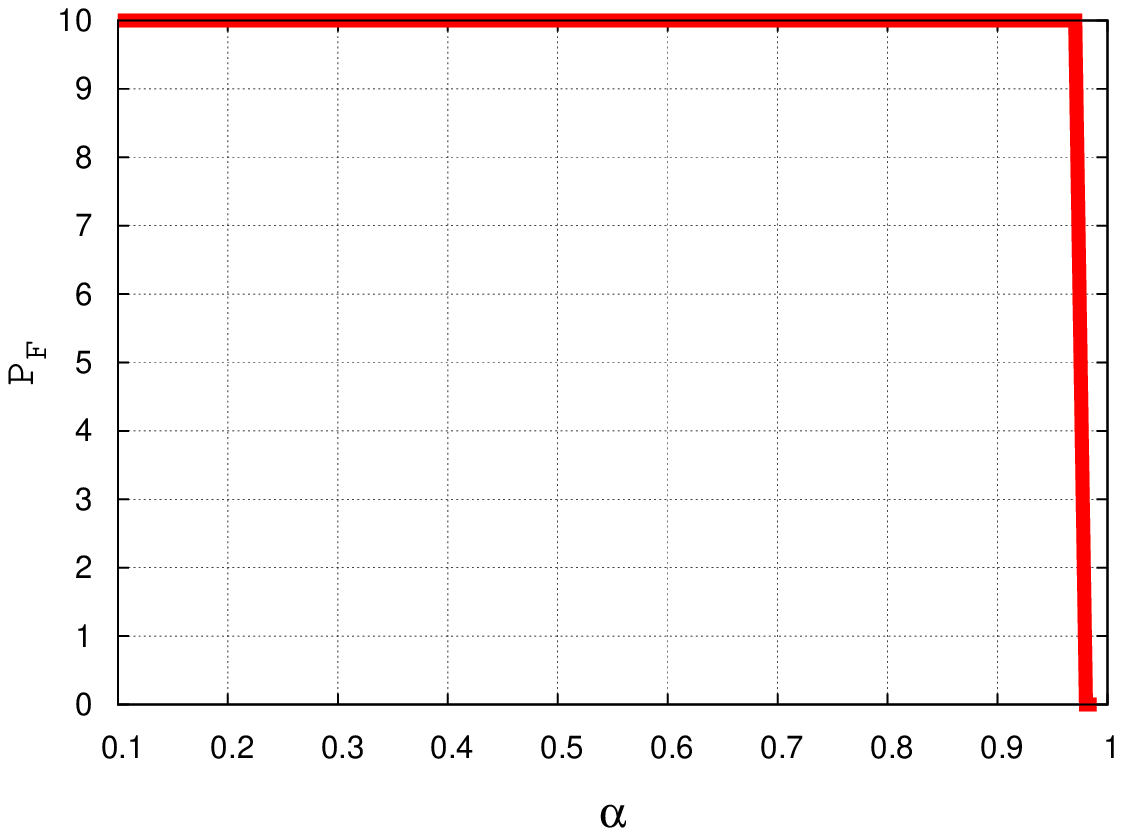}&
    \includegraphics[totalheight=.2 \textheight, width=.5\textwidth]{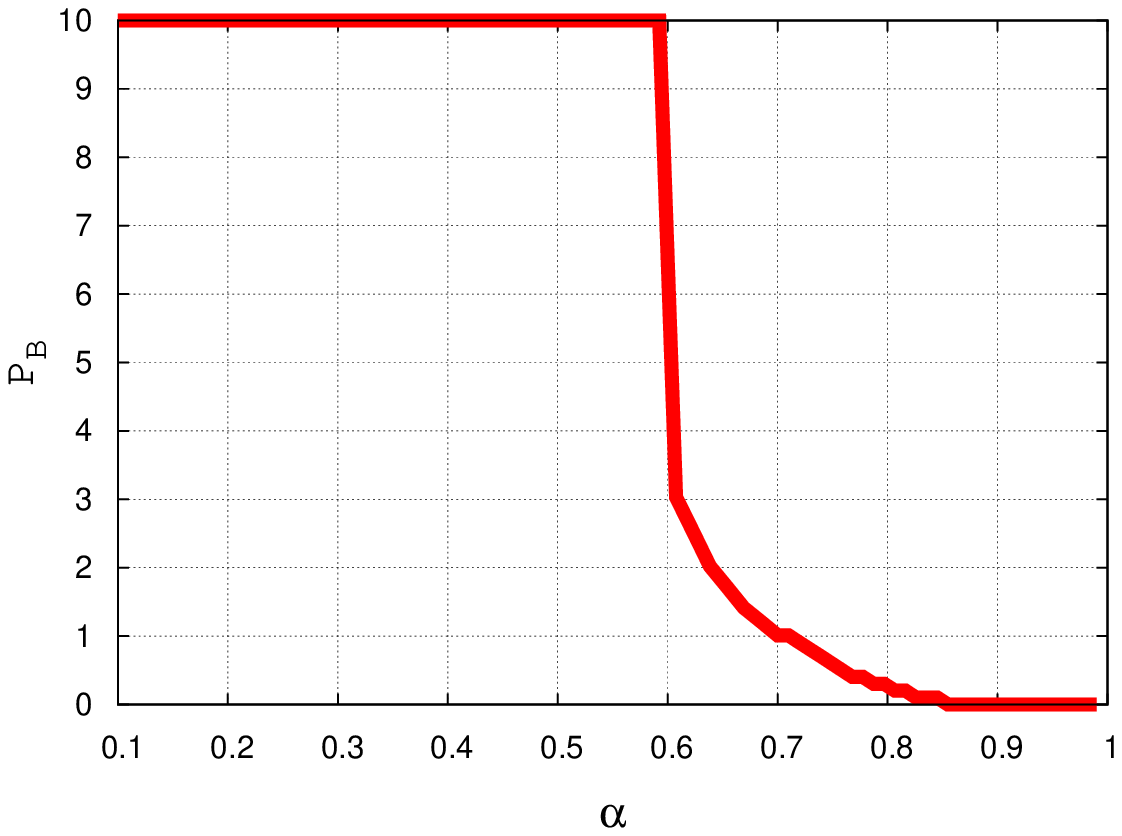}\\
    \end{tabular}
    \caption{Perfect sensing power and transmission time results for channel \textbf{B}.}
    \label{fig:perfect_channelB}
    \end{figure*}
\begin{figure*}
    \begin{tabular}{cc}
        \includegraphics[totalheight=.2\textheight,width=.5\textwidth]{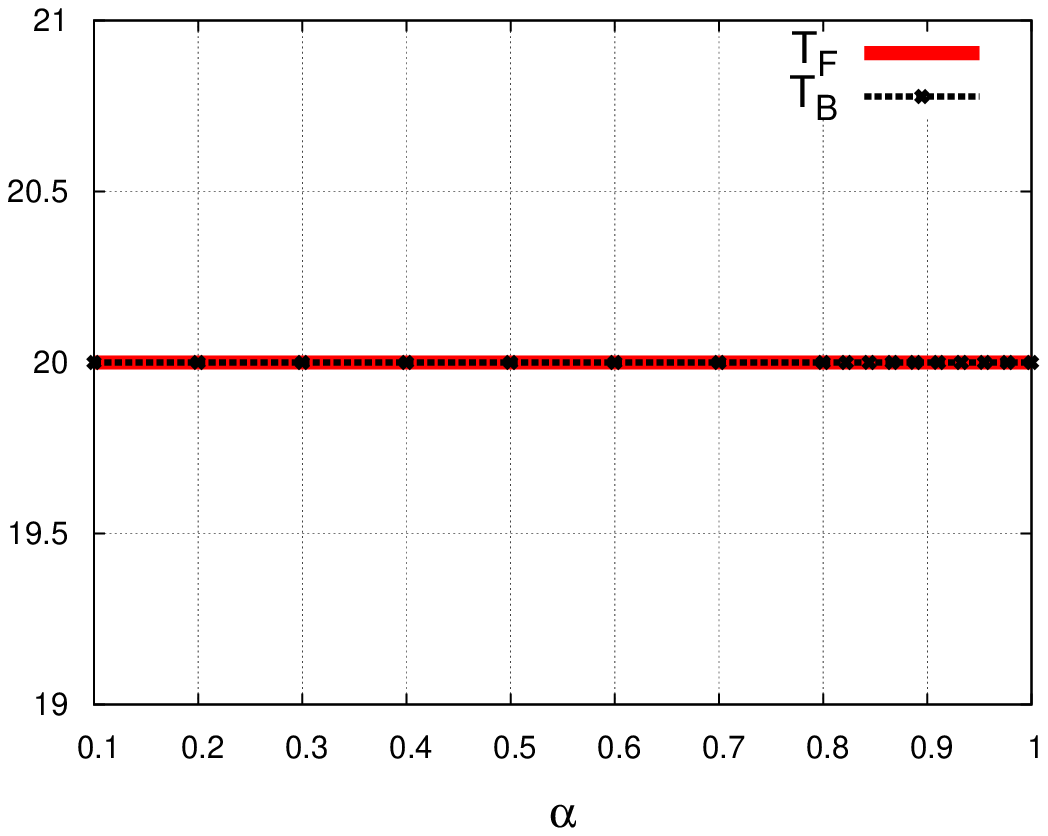}&
    \includegraphics[totalheight=.2 \textheight, width=.5\textwidth]{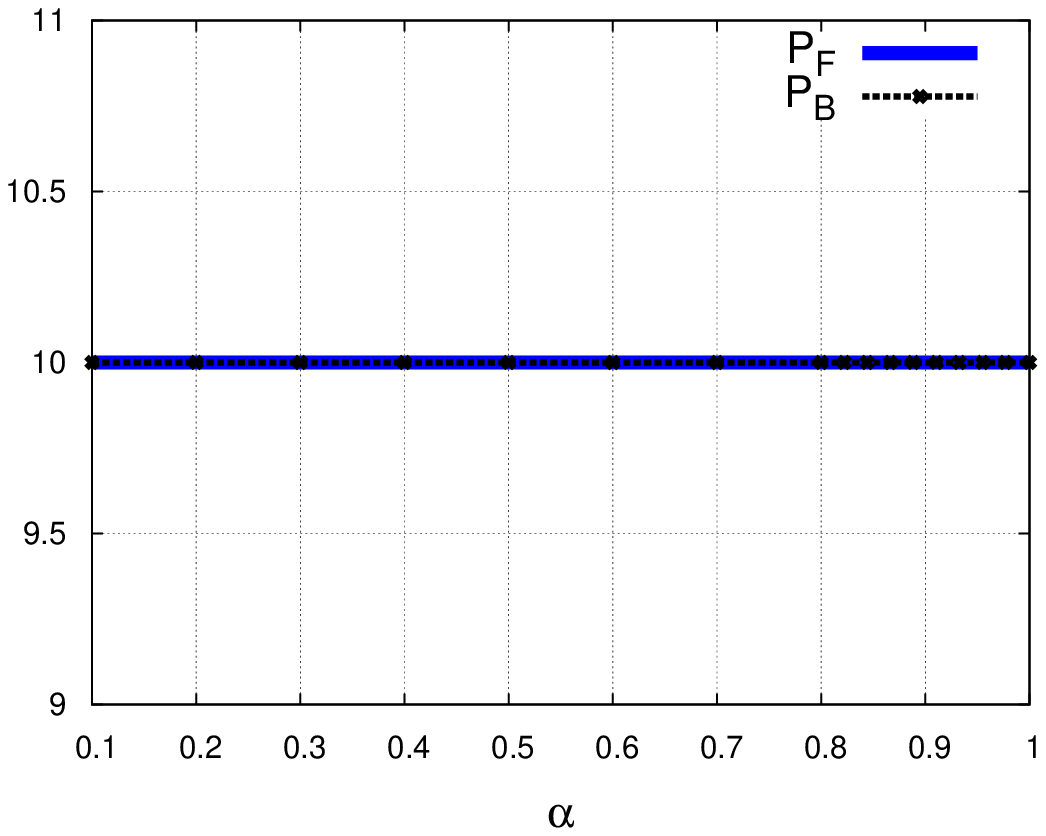}\\
    \end{tabular}
    \caption{Perfect sensing power and transmission time results for $g_{sp}$=0.002.}
    \label{fig:perfect_small gsp}
\end{figure*}

The weighted sum throughput versus $\alpha$ is shown in Figure \ref{fig:rate_alpha_perfect} for channels \textbf{A} and \textbf{B}. It is clear from the figure that as the gain $\overline{g}_{\rm sp}$ increases, the level of interference at the primary receiver increases leading to lower data rates. The optimal transmission power and time parameters for channel \textbf{A} are given in Figure \ref{fig:parameter_d_perfect}. For small $\alpha$ value, which corresponds to giving more importance to the secondary throughput, the secondary transmitter emits $P_{\rm max}$ whether the channel is sensed to be free or busy. The transmission time for both sensing outcomes are the maximum possible. Recall that this maximum is artificial and is imposed by the exhaustive search solution. In fact, for $\alpha$ approaching zero, the secondary transmitter sends with $P_{\rm max}$ continuously without the need to sense the channel again. If the optimal $P_{\rm F}=P_{\rm B}$, then sensing becomes superfluous because the exact same power would be used regardless of the sensing outcome. As $\alpha$ increases, the power transmitted when the channel is sensed to be busy is reduced below $P_{\rm max}$. In addition, the transmission times are reduced for more frequent checking of primary activity. As $\alpha$ approaches unity, the secondary transmitter is turned off and the channel is not sensed. Figure \ref{fig:perfect_channelB} gives the optimal transmission parameters for channel \textbf{B}. It is evident from the figure that as the level of interference from secondary transmitter to primary receiver is decreased, $P_{\rm B}$ becomes lower than $P_{\rm max}$ at a higher $\alpha$ compared to \textbf{A}. If we make $\overline{g}_{\rm sp}=0.002$, the secondary transmits all time with maximum power regardless of the sensing outcome. This is shown in Figure \ref{fig:perfect_small gsp}.
\subsection{Soft Sensing}
\label{sim_soft}
\begin{figure*}[htpb]
    \centering
    \begin{tabular}{cc}
        \includegraphics[totalheight=.2\textheight,width=.5\textwidth]{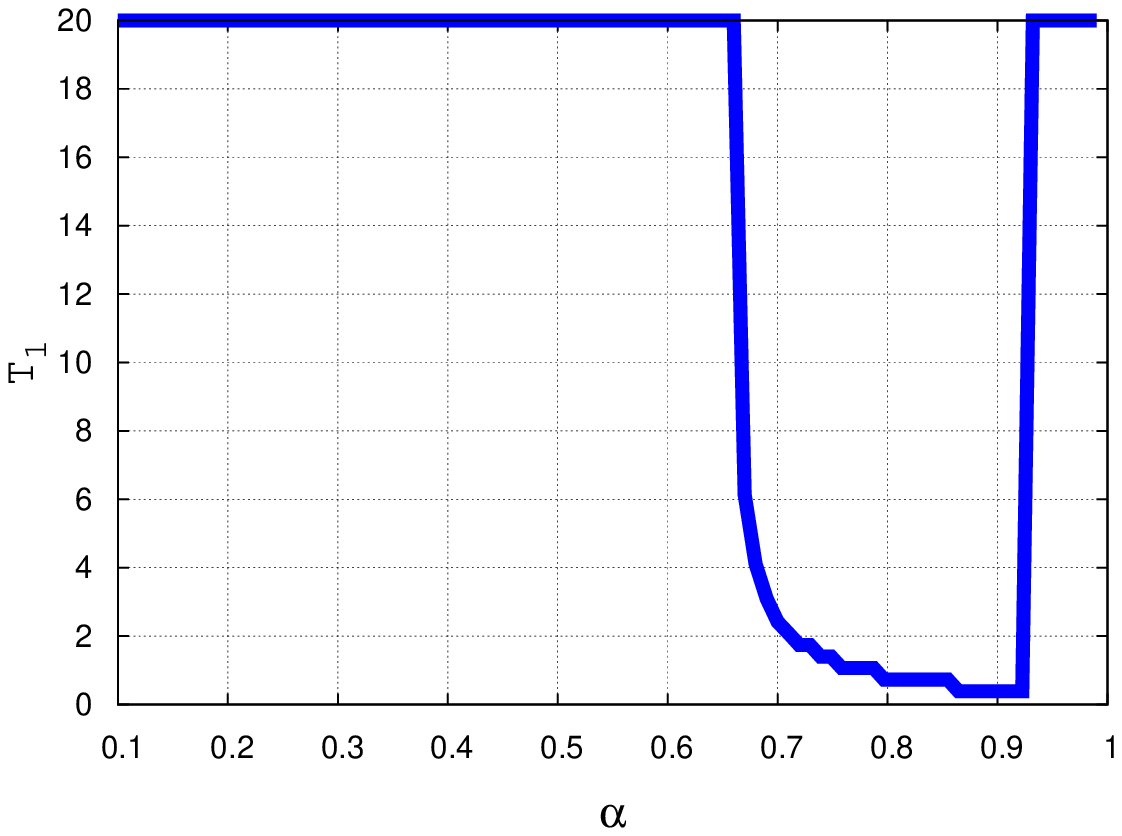}&
        \includegraphics[totalheight=.2 \textheight, width=.5\textwidth]{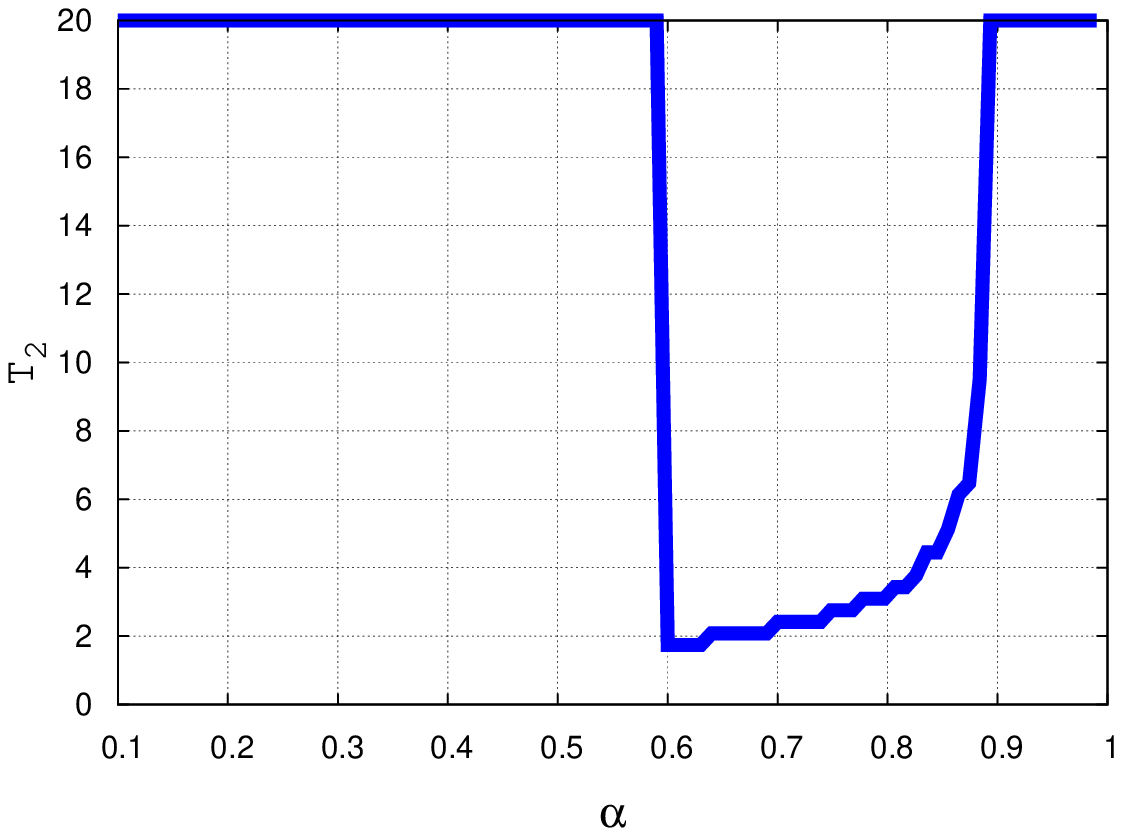}\\
    \includegraphics[totalheight=.2\textheight, width=.5\textwidth]{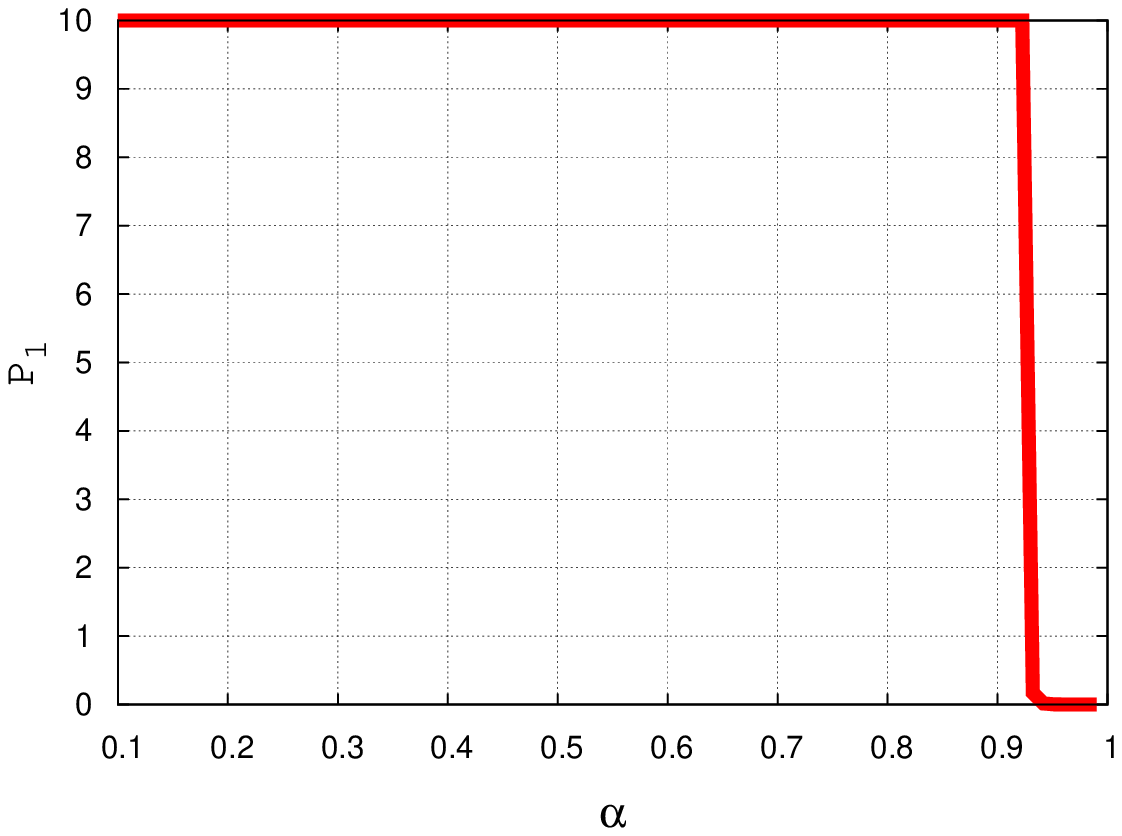}&
    \includegraphics[totalheight=.2 \textheight, width=.5\textwidth]{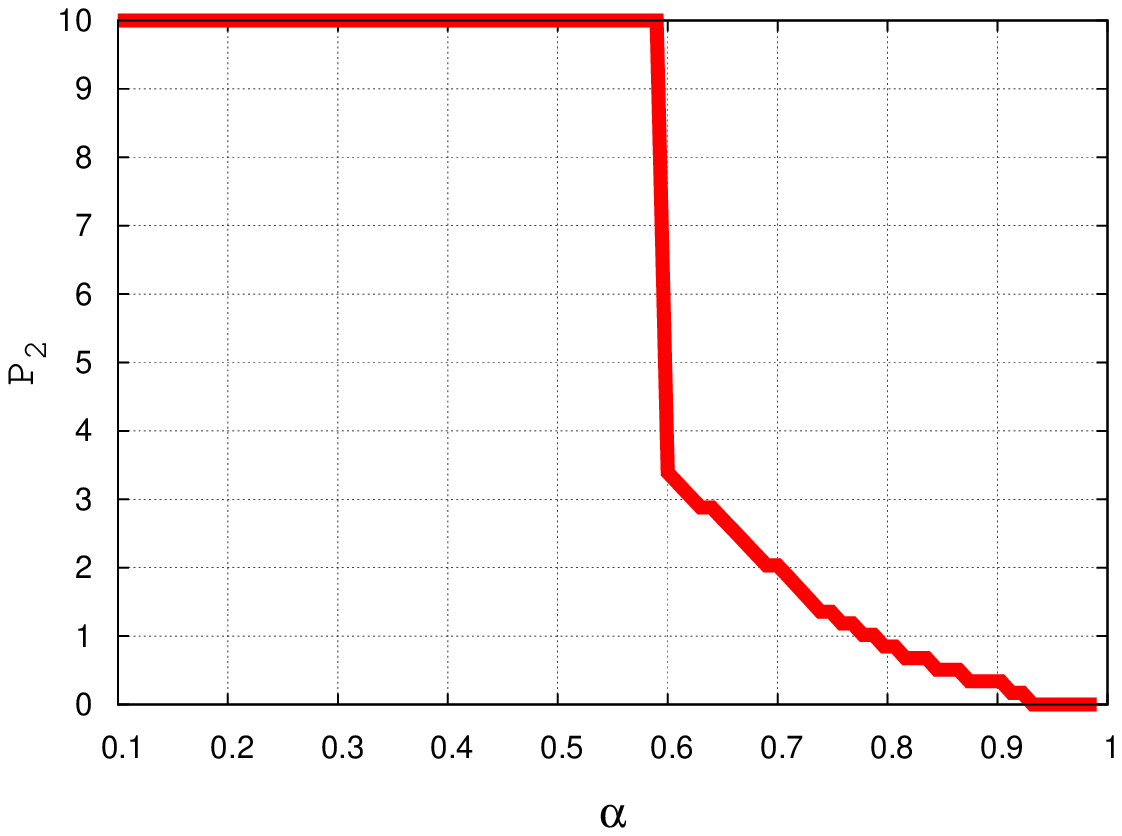}\\
    \end{tabular}
    \caption{Soft sensing power and transmission time results for channel \textbf{B}.}
    \label{fig:soft_channelB}
\end{figure*}

\begin{figure}[htpb]
    \centering
\includegraphics[width=.5\textwidth,height=7cm]{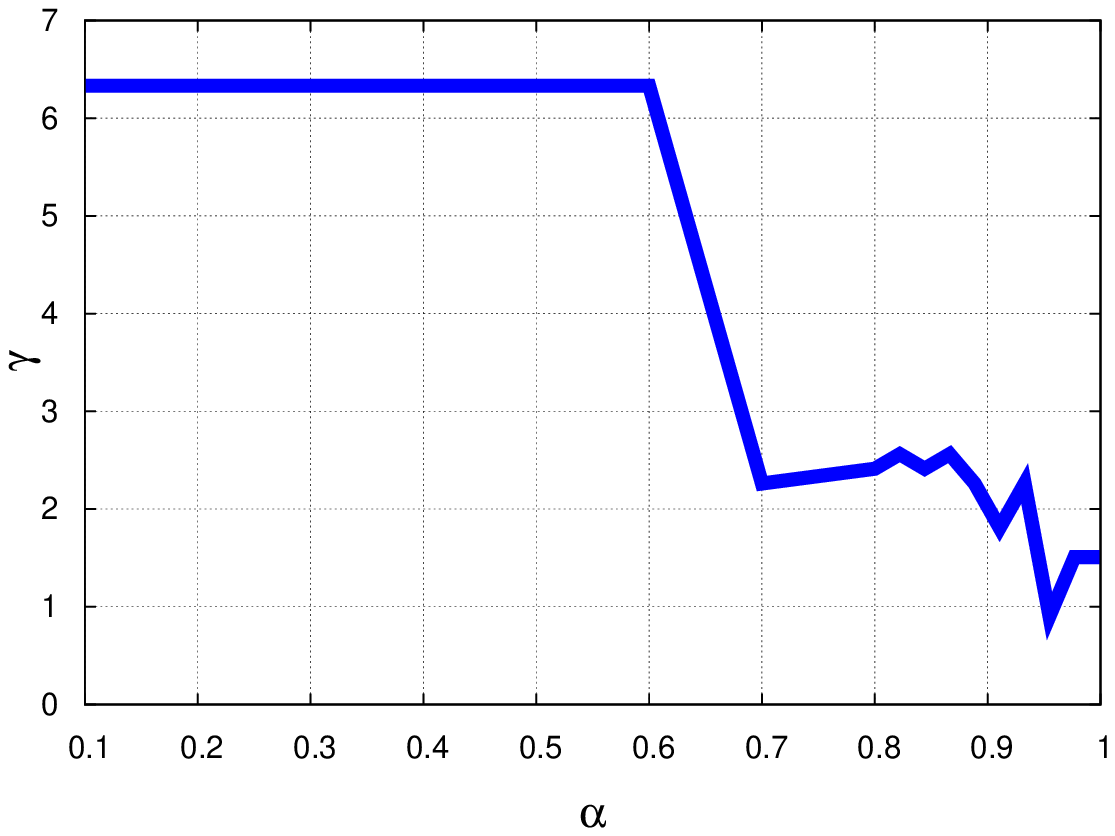}
\caption{Soft sensing optimal threshold for channel \textbf{B}.}
    \label{fig:soft_thr_channelB}
\end{figure}

In the soft sensing case, the optimization parameters are $2\left(S+1\right)$ transmission powers and times corresponding to each quantization level. There are also $S$ thresholds defining the boundaries of the quantization levels. Hence, the total number of parameters is $3S+2$. The conditional distributions of the sensing metric $\gamma$ used in the simulations are $f_{\circ}\left(\gamma\right)=\exp\left(-\gamma\right)$ and $f_{\rm 1}\left(\gamma\right)=\exp\left(-\left[\gamma+\gamma_{\circ}\right]\right)\ I_{\circ}\left(2\sqrt{\gamma \gamma_{\circ}}\right)$, where $I_{\circ}$ is the zero order modified Bessel function and $\gamma_{\circ}$ is a parameter related to the mean value of $f_{\rm 1}\left(\gamma\right)$. We present here the results for one and two thresholds. The case of one threshold corresponds to the imperfect sensing case where the primary is assumed to be active when $\gamma$ exceeds some threshold and inactive otherwise. The false alarm probability is given by $\epsilon_{\rm 2}$, whereas the miss detection probability is $\vartheta_{\rm 1}$.
Figures \ref{fig:soft_channelB} and \ref{fig:soft_thr_channelB} give  the optimal parameters as a function of $\alpha$ and for $\gamma_{\circ}=3$. As is evident from the figure, the optimal threshold decreases with
\begin{figure}[htpb]
    \centering
       \psfrag{R}[][]{$\rm {\alpha\ \overline{R}_p + (1-\alpha)\ \overline{R}_s}$} 
    \epsfig{file=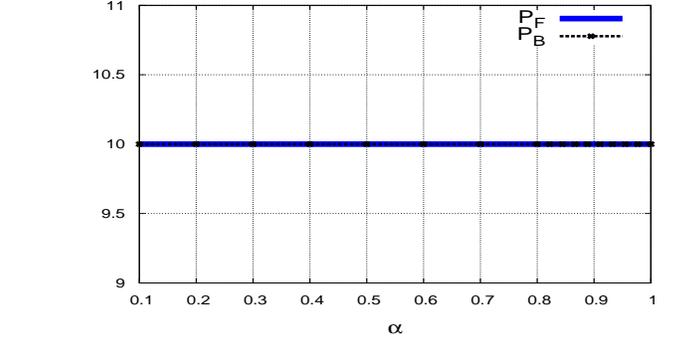, width=0.5\textwidth,height=7cm}
    \caption{Soft sensing weighted sum throughput for one threshold with different $\gamma_0$ versus $\alpha$ for channel \textbf{B}.}
    \label{fig:soft_2thre_gamma3_and_10_c}
\end{figure}
$\alpha$. Under the imperfect sensing interpretation of the one threshold case, this means that as $\alpha$ increases putting more emphasis on the primary rate, the required false alarm probability is increased while the miss detection probability is decreased to reduce the chance of collision with the primary user. The effect of different values for $\gamma_{\circ}$ is given in Figure \ref{fig:soft_2thre_gamma3_and_10_c} where the weighted sum rate is higher for $\gamma_{\circ}=10$ than for $\gamma_{\circ}=3$. This is attributed to the increased distance between the distributions $f_{\circ}\left(\gamma\right)$ and $f_{\rm 1}\left(\gamma\right)$, thereby lowering the false alarm and miss detection probabilities.
\begin{figure}[htpb]
    \centering
      \psfrag{R}[][]{$\rm {\alpha\ \overline{R}_p + (1-\alpha)\ \overline{R}_s}$} 
      \psfrag{One threshold}[][]{One threshold}
      \psfrag{Two thresholds}[][]{Two thresholds}
      \psfrag{Perfect sensing}[][]{Perfect sensing}
    \epsfig{file=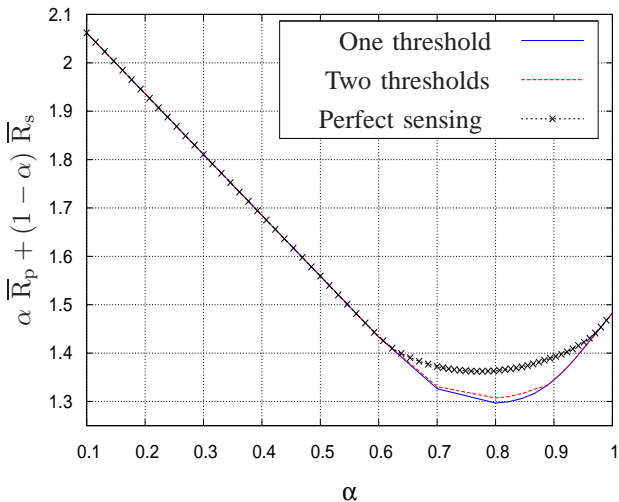, width=0.5\textwidth,height=7cm}
\caption{Soft sensing weighted sum throughput versus $\alpha$  using one and two thresholds for channel \textbf{B}. The results from perfect sensing is provided for comparison.}
    \label{fig:one_two_threshod}
\end{figure}
Figure \ref{fig:one_two_threshod} shows the weighted sum throughput using one and two thresholds for channel \textbf{B} and $\gamma_{\circ}=3$. There is a range of $\alpha$ values for which the two-threshold scheme improves very slightly the weighted sum rates. 
\section{Conclusion}
\label{sec:Conc}

We have investigated the problem of specifying transmission power
and duration in an underlay unslotted cognitive radio network,
where the primary transmission duration  follows an exponential
distribution. We used an upperbound for the secondary throughput,
and obtained, numerically, the optimal secondary transmission
power and duration that maximize a weighted sum of the primary and
secondary throughputs. Note that, at the particular values
obtained,  the solutions obtained from our optimization problem,
are the same that would be obtained from a constrained
optimization problem where one seeks to maximize the secondary
throughput while constraining the primary throughput to be above a
certain value. Our results also showed that an increase in the
overall weighted throughput can be obtained by allowing the
secondary to transmit even when the channel is found to be busy.
We extended our formulation to the soft sensing case where the
decision of the secondary transmission power and duration depends
on the quantized value of the sensing metric, rather than on the
binary decision of whether the channel is free or not. However,
our preliminary results show that the gain of using this scheme,
and for the range of parameters we have simulated, are minimal.

\section{Appendix}
\label{sec:appendix}

We provide here the evaluation of (\ref{P_outage}), (\ref{sec_cap_no_primary}), and (\ref{sec_cap_with_primary}) for exponential channel gains. The outage probability (\ref{P_outage}) can be written as

\begin{eqnarray}
P_{\circ}\left(p\right) &=& {\rm Pr}\left\{r_{\circ} > \log\left(1+\frac{ag_{\rm pp}}{bg_{\rm sp}+1}\right)\right\} \nonumber\\
&=& {\rm Pr}\left\{ag_{\rm pp}-bcg_{\rm sp} < c \right\}
\end{eqnarray}
where $a=P_{\rm p}/\sigma_{\rm p}^2$, $b=p/\sigma_{\rm p}^2$, and
$c=\exp\left(r_{\circ}\right)-1$. Assuming that $g_{\rm pp}$ and
$g_{\rm sp}$ are independent and exponentially distributed with
means $\overline{g}_{\rm pp}$ and $\overline{g}_{\rm sp}$, the
outage probability becomes
\begin{eqnarray}
P_{\circ}\left(p\right)&=&
\int^{\infty}_{0}\int^{\frac{c}{a}\left(1+bg_{\rm
sp}\right)}_{0}\frac{1}{\overline{g}_{\rm pp}} \exp(-\frac{g_{\rm
pp}}{\overline{g}_{\rm pp}})\times \nonumber\\ &&\
\frac{1}{\overline{g}_{\rm sp}}\exp(-\frac{g_{\rm sp}}{\overline
{g}_{\rm sp}})dg_{\rm pp}dg_{\rm sp} \nonumber\\ &=&\
1-\frac{P_{\rm p}\overline{g}_{\rm pp}}{P_{\rm p}\overline{g}_{\rm
pp}+pc\overline{g}_{\rm sp}} \exp\left(-\frac{c\sigma^2_{\rm
p}}{P_{\rm p}\overline{g}_{\rm pp}}\right)
\end{eqnarray}

Assuming an exponential distribution for $g_{\rm ss}$ with mean
$\overline{g}_{\rm ss}$, (\ref{sec_cap_no_primary}) becomes
\begin{eqnarray}
C_{\circ}\left(p\right)=\int_{0}^{\infty}\log\left(1+\frac{pg_{\rm
ss}}{\sigma^2_{\rm s}}\right)\frac{1}{\overline{g}_{\rm
ss}}\exp\left(-\frac{g_{\rm ss}}{\overline{g}_{\rm
ss}}\right)dg_{\rm ss}
\end{eqnarray}
Defining $\Psi\left(x\right)=\int_{x}^{\infty}\exp(-\mu)/\mu\ d\mu$, it is straightforward to show that
\begin{eqnarray}
C_{\circ}\left(p\right)=\exp\left(\frac{\sigma^2_{\rm
s}}{p\overline{g}_{\rm ss}}\right) \Psi \left(\frac{\sigma^2_{\rm
s}}{p\overline{g}_{\rm ss}}\right)
\end{eqnarray}

Assuming that $g_{\rm ss}$ and $g_{\rm ps}$ are independent and
have means $\overline{g}_{\rm ss}$ and $\overline{g}_{\rm ps}$,
respectively, (\ref{sec_cap_with_primary}) can be expressed as
\begin{eqnarray}
\label{mean_log(g1_and_g2)}
C_{\rm 1}\left(p\right)&=& \int_{0}^{\infty}\int_{0}^{\infty}\log\left(1+\frac{pg_{\rm ss}}{P_{\rm p}g_{\rm ps}+\sigma^2_{\rm s}}\right)) \times \nonumber\\
&& \frac{1}{\overline{g}_{\rm ss}}\exp(-\frac{g_{\rm
ss}}{\overline{g}_{\rm ss}})
 \frac{1}{\overline{g}_{\rm ps}}\exp(\frac{g_{\rm ps}}{\overline{g}_{\rm ps}})dg_{\rm ss}dg_{\rm ps} \nonumber\\
 &=& C_{\rm 1a}-C_{\rm 1b}
\end{eqnarray}
where
\begin{eqnarray}
C_{\rm 1a}&=& \int_{0}^{\infty}\int_{0}^{\infty}\log\left(1+\frac{P_{\rm p}}{\sigma^2_{\rm s}}g_{\rm ps}+\frac{p}{\sigma^2_{\rm s}}g_{\rm ss}\right) \times \nonumber\\
&& \frac{1}{\overline{g}_{\rm ss}}\exp(-\frac{g_{\rm
ss}}{\overline{g}_{\rm ss}})
 \frac{1}{\overline{g}_{\rm ps}}\exp(\frac{g_{\rm ps}}{\overline{g}_{\rm ps}})dg_{\rm ss}dg_{\rm ps}
\end{eqnarray}
\begin{eqnarray}
C_{\rm 1b}&=&\int_{0}^{\infty}\log\left(1+\frac{P_{\rm p}g_{\rm ps}}{\sigma^2_{\rm s}}\right)\frac{1}{\overline{g}_{\rm ps}}\exp\left(-\frac{g_{\rm ps}}{\overline{g}_{\rm ps}}\right)dg_{\rm ps} \nonumber\\
&=& \exp\left(\frac{\sigma^2_{\rm s}}{P_{\rm p}\overline{g}_{\rm
ps}}\right) \Psi \left(\frac{\sigma^2_{\rm s}}{P_{\rm
p}\overline{g}_{\rm ps}}\right)
\end{eqnarray}
We find $C_{\rm 1a}$ by rewriting it as
\begin{eqnarray}
C_{\rm 1a}&=& \int_{0}^{\infty}\log\left(1+z\right)f\left(z\right)dz
\end{eqnarray}
where $z=\frac{P_{\rm p}}{\sigma^2_{\rm s}}g_{\rm ps}+\frac{p}{\sigma^2_{\rm s}}g_{\rm ss}$ and $f\left(z\right)$ is the pdf of $z$. If $x=\frac{P_{\rm p}}{\sigma^2_{\rm s}}g_{\rm ps}$ and $y=\frac{p}{\sigma^2_{\rm s}}g_{\rm ss}$, then $x$ and $y$ are independent and have the exponential distributions
\begin{equation}
f_{\rm X}\left(x\right)=\frac{\sigma^2_{\rm s}}{P_{\rm
p}\overline{g}_{\rm ps}}\exp\left(-\frac{\sigma^2_{\rm s}x}{P_{\rm
p}\overline{g}_{\rm ps}}\right),\, x\geq 0 \nonumber
\end{equation}
and
\begin{equation}
f_{\rm Y}\left(y\right)=\frac{\sigma^2_{\rm s}}{p\overline{g}_{\rm
ss}}\exp\left(-\frac{\sigma^2_{\rm s}y}{p\overline{g}_{\rm
ss}}\right),\, y\geq 0 \nonumber
\end{equation}
respectively. The pdf $f\left(z\right)$ is the convolution of $f_{\rm X}\left(x\right)$ and $f_{\rm Y}\left(y\right)$:
\begin{eqnarray}
f\left(z\right)&=&\int_{0}^{z}f_{\rm X}\left(x\right)f_{\rm Y}\left(z-x\right)dx \nonumber\\
&=& \sigma^2_{\rm s}\frac{\exp\left(-\frac{\sigma^2_{\rm
s}z}{p\overline{g}_{\rm ss}}\right)-\exp\left(-\frac{\sigma^2_{\rm
s}z}{P_{\rm p}\overline{g}_{\rm ps}}\right)}{p\overline{g}_{\rm
ss}-P_{\rm p}\overline{g}_{\rm ps}}
\end{eqnarray}
Note that in the case $p\overline{g}_{\rm ss}=P_{\rm
p}\overline{g}_{\rm ps}=w$, we can use L'H${\rm \hat{o}}$pital's
rule to get
\begin{eqnarray}
f\left(z\right)&=& \frac{\sigma^4_{\rm s}z}{w^2}\exp\left(-\frac{\sigma^2_{\rm s}z}{w}\right)
\end{eqnarray}
It can then be shown that when $p\overline{g}_{\rm ss} \neq P_{\rm
p}\overline{g}_{\rm ps}$
\begin{eqnarray}
C_{\rm 1a}&=& \frac{1}{p\overline{g}_{\rm ss}-P_{\rm p}\overline{g}_{\rm ps}}{\Big [}p\overline{g}_{\rm ss}\exp\left(\frac{\sigma^2_{\rm s}}{p\overline{g}_{\rm ss}}\right) \Psi \left(\frac{\sigma^2_{\rm s}}{p\overline{g}_{\rm ss}}\right)- \nonumber\\
&& P_{\rm p}\overline{g}_{\rm ps}\exp\left(\frac{\sigma^2_{\rm
s}}{P_{\rm p}\overline{g}_{\rm ps}}\right) \Psi
\left(\frac{\sigma^2_{\rm s}}{P_{\rm p}\overline{g}_{\rm
ps}}\right){\Big ]}
\end{eqnarray}
Therefore,
\begin{eqnarray}
C_{\rm 1}\left(p\right)&=& \frac{p\overline{g}_{\rm ss}}{p\overline{g}_{\rm ss}-P_{\rm p}\overline{g}_{\rm ps}}{\Big [}\exp\left(\frac{\sigma^2_{\rm s}}{p\overline{g}_{\rm ss}}\right) \Psi \left(\frac{\sigma^2_{\rm s}}{p\overline{g}_{\rm ss}}\right)- \nonumber\\
&& \exp\left(\frac{\sigma^2_{\rm s}}{P_{\rm p}\overline{g}_{\rm
ps}}\right) \Psi \left(\frac{\sigma^2_{\rm s}}{P_{\rm
p}\overline{g}_{\rm ps}}\right){\Big ]}
\end{eqnarray}
In the case $p\overline{g}_{\rm ss}=P_{\rm p}\overline{g}_{\rm
ps}=w$,
\begin{eqnarray}
C_{\rm 1a}=\int_{0}^{\infty}\frac{\sigma^4_{\rm s}z}{w^2}\exp\left(-\frac{\sigma^2_{\rm s}z}{w}\right)\log\left(1+z\right)dz \nonumber
\end{eqnarray}
Integrating by parts we obtain
\begin{eqnarray}
C_{\rm 1a}=1+\left(1-\frac{\sigma^2_{\rm s}}{w}\right)\exp\left(\frac{\sigma^2_{\rm s}}{w}\right) \Psi \left(\frac{\sigma^2_{\rm s}}{w}\right) \nonumber
\end{eqnarray}
and, hence, when $p\overline{g}_{\rm ss}=P_{\rm
p}\overline{g}_{\rm ps}$
\begin{eqnarray}
C_{\rm 1}\left(p\right)=1-\frac{\sigma^2_{\rm
s}}{p\overline{g}_{\rm ss}}\exp\left(\frac{\sigma^2_{\rm
s}}{p\overline{g}_{\rm ss}}\right) \Psi \left(\frac{\sigma^2_{\rm
s}}{p\overline{g}_{\rm ss}}\right)
\end{eqnarray}

\end{document}